\documentclass[aps,twocolumn,a4paper,10pt]{revtex4}
\usepackage{mathrsfs}
\usepackage{graphicx}
\usepackage{latexsym}
\usepackage{amsmath}
\usepackage{amssymb}
\usepackage{textcomp}
\usepackage{amsbsy}
\usepackage{graphics}
\begin{document}
\begin{flushright}
HRI-RECAPP-2014-022
\end{flushright}

\title{Why has spacetime torsion such  negligible  effect on our universe? }
\preprint{HRI-RECAPP-2014-022}
\author{Ashmita Das${}^\dagger$, Biswarup Mukhopadhyaya${}^{\#}$ and Soumitra SenGupta${}^\dagger$}
\email{tpad@iacs.res.in,biswarup@hri.res.in,tpssg@iacs.res.in}
\affiliation{${}^\dagger$Department of Theoretical Physics, Indian association For The Cultivation
 Of science, Kolkata 700032, India}
\affiliation{${}^{\#}$Regional Centre for Accelerator-based Particle Physics,
Harish-Chandra Research Institute, Chhatnag Road, Jhusi, Allahabad -
211 019, India.}
% \affiliation{${}^{*}$Department of Theoretical Physics, Indian Association for the
% Cultivation of Science
% \\ 2A and 2B Raja S.C. Mallick Road, Jadavpur, Kolkata 700 032, India.}
%\date{\today}

\begin{abstract}
We attempt an answer to the question as to why the evolution of 
four-dimensional universe is governed by spacetime curvature but
not torsion.  An answer is found if there is an additional
compact spacelike dimension with a warped geometry, with torsion caused
by a Kalb-Ramond (KR) antisymmetric tensor field in the bulk. Starting from 
a Randall-Sundrum type of warped extra dimension, and including the
inevitable back reaction ensuing from the radius stabilization mechanism,
we show that there is always an extra exponential suppression of the KR field
on the four-dimensional projection that constitutes our visible universe.
The back reaction is found to facilitate the process of such suppression. 
\end{abstract}

\maketitle

An enigmatic feature of our universe, as evinced from cosmological
observations, is that its evolution is controlled by one type of
geometrical deformation only, namely, curvature. It is somewhat
surprising that we notice practically no effect of another type of
deformation, namely, torsion. The simplest extension of the general
theory of relativity incorporating torsion is the Einstein-Cartan
theory, involving the antisymmetric rank-3 torsion
$H_{\mu\nu\lambda}$. It can be argued from dimensional consideration
that the coupling of $H_{\mu\nu\lambda}$ to matter should be $\sim
1/M_P$ where $M_P = 1/\sqrt{G}$ is the scale answering to any theory
of gravity, namely, the Planck scale.  There have been several
experimental searches to verify Einstein's theory of gravity without any presence of torsion in spacetime
geometry.  An example is the Gravity Probe B experiment which was
designed to estimate the precession of a
gyroscope to observe any signature beyond the predictions of Einstein's gravity 
\cite{everitt}. However, all such probes, within the limit of their
experimental precision, have consistently produced negative results and
thereby disfavoured the presence of torsion in the spacetime geometry
of our (3+1) dimensional visible universe \cite{hehl,guth,lanmerzahl}.
The question, therefore, is: why does torsion leave such negligible footrpint
on the universe while curvature controls evolution almost entirely?
Neither pure Einstein gravity nor Einstein-Cartan models can answer
this.

A rather spectacular solution is offered in terms of theories
involving extra compact spacelike dimensions, with warped geometry for
the compact dimension(s), such as in Randall-Sundrum (RS) \cite{RS1} scenarios.
This kind of a scenario postulates gravity in the five-dimensional
`bulk', whereas our four-dimensional universe is confined to one of the two 3-branes
located at the two orbifold fixed points along the compact dimension. 
  
It has been shown already that, if there is a rank-2 antisymmetric
tensor field $B_{\mu\nu}$, popularly called the Kalb-Ramond (KR)
field, then the rank 3 field strength tensor $H_{\mu\nu\lambda}$ corresponding to 
the gauge field $B_{\mu\nu}$ can be equated with the source of 
spacetime torsion using the equation of motion \cite{pmssg}. 
In an RS-like scenario with, say, one extra
dimension, both the graviton and $B_{\mu\nu}$ propagate in the bulk.
This can also find justification in string theory where the graviton
and the KR field are both excitations of closed strings. In such a
case, the compactification procedure causes the KR field to be diluted
compared to the graviton by an exponential warp factor on the 3-brane
where the standard model (SM) fields are localised \cite{ssgkr}.

The robustness of such a claim, however, awaits a crucial demonstration.
The warped RS \cite{RS1} geometry in its original form is intrinsically unstable. A popular way
of stabilising the radius of the compact dimension is the Goldberger-Wise (GW) \cite{GW1}
mechanism, postulating a bulk scalar field with different vacuum expectation values
(VEV) at the two 3-branes that reside at the orbifold fixed points of
$S_1/Z_2$ compactification. This mechanism, however, generates a bulk energy
density which modifies the warped geometry itself via back-reaction. Though 
this had been neglected in the initial GW proposal, its various implications,
including the modifications of the warp factor, 
have been subsequently investigated \cite{kribs,wolfe}. We show here that such back reaction
retains the dilution of the zero-mode of the KR field (which
is the instrument of torsion) on the SM brane, thus putting this
explanation of vanishing torsion in our universe on a firm basis.  

The minimal RS scenario is based on the five-dimensional metric
\begin{equation}
ds^2=e^{-2ky}\eta_{\mu\nu}dx^{\mu}dx^{\nu}-dy^{2}\label{mwarpfactor1}
\end{equation}
\noindent
where $k (\approx M_P)$ is related to the bulk cosmological constant,
The brane on which the SM fields reside is located at $y = r_0$, while
the `Planck brane' is at $y = 0$. Achieving the correct hierarchy
between the Planck and TeV scales requires $k r_0 \approx 36$.  The
stability of this desired $r_0$, however, is not assured. The GW
mechanism aims to achieve such stabilisation by postulating a bulk
scalar field with an appropriate potential for the scalar field in the
bulk.

We wish to see how torsion, equated with the KR field, fares on the SM brane
in the above geometry. For that, however,   
it is imperative to examine the consequence of this bulk scalar energy density
on the geometry, or in other words, the effects of `back reaction'.
With this in view, we use
the modified warp factor to determine the zero-mode of the KR
field strength which corresponds to the space-time torsion in our
brane. Then we examine its coupling with the SM matter fields localised
on the visible brane.

The five-dimensional action used here is \cite{kribs}
\begin{eqnarray}
S&=&-M^3\int d^{5}x\sqrt{g}R^{(5)}+\int d^{5}x\sqrt{g}(\frac{1}{2}\nabla \phi \nabla \phi-V(\phi))\nonumber\\
&-&\int d^4x\sqrt{g_4}\lambda_{P}(\phi)-\int d^4x\sqrt{g_4}\lambda_{T}(\phi)\nonumber\\
&+& \int d^{5}x H_{MNL}H^{MNL}\label{actionbr}
\end{eqnarray}
Where $R^{(5)}$ is the five dimensional Ricci scalar, $\phi$ is the bulk scalar field and $V(\phi)$ is
the bulk potential term for the scalar field $\phi$, $g_4$ is the 
induced metric on the brane and $\lambda_P$, $\lambda_T$ are the boundary potentials
on the Planck and TeV branes respectively due to the  bulk scalar field. 
The generalised metric is written as
\begin{equation}
ds^2=e^{-2A(y)}\eta_{\mu\nu}dx^{\mu}dx^{\nu}-dy^{2}\label{mwarpfactor2}
\end{equation}
preserve 4-D Lorentz invariance where $y$ symbolises as extra dimensional coordinate
and $e^{-A(y)}$ is the new modified warp factor \cite{kribs}.\\
We initially consider the back-reaction of only the bulk scalar field. Later we will comment
on the effects of similar back-reaction of the bulk KR field also. Under this assumption,  
the five dimensional Einstein equations are given as, 
\begin{equation}
R_{AB}=\kappa^{2}(T_{AB}-\frac{1}{3}g_{AB}g^{CD}T_{CD})\label{eeq2}
\end{equation}
Here $\kappa$ is the five dimensional Newton's constant and related to 
five dimensional Planck mass $M$ by $\kappa^2=1/2M^3$.\\
As shown in \cite{kribs}, various components of the 5-D Einstein equations for this metric ansatz
are:
\begin{equation}
 4A'^2-A''=-\frac{2\kappa^2}{3}V(\phi)-\frac{\kappa^2}{3}
\sum_{i}\lambda_i(\phi)\delta(y-y_i)\label{fieldeq1}
\end{equation}
\begin{equation}
 A'^2=\frac{\kappa^2\phi^{'2}}{12}-\frac{\kappa^2}{6}V(\phi)\label{fieldeq2}
\end{equation}
\begin{equation}
 \phi_0^{''}=4A'\phi^{'}+\frac{\partial V(\phi)}{\partial \phi}+\sum_i
\frac{\partial \lambda_i(\phi)}{\partial \phi}\delta(y-y_i)\label{scfieldeq1}
\end{equation}
Here prime and $\partial_\mu$ denotes the derivatives with respect to $y$ and 
4-D space time coordinate i.e $x^\mu$ respectively.
Thus the energy density of the bulk scalar field interacts gravitationally with the 
bulk spacetime to modify the effective energy-momentum tensor.
To solve the Einstein's
equations in this spacetime geometry one considers a class of potential for
 the bulk scalar field $\phi$ \cite{kribs,wolfe} inspired from 
5-dimensional gauged supergravity \cite{freedman} as,
\begin{equation}
 V(\phi)=\frac{1}{8}(\frac{\partial W(\phi)}{\partial \phi})^2-\frac{\kappa^2}{6}
W(\phi)^2\label{potform1}
\end{equation}
Where,
\begin{equation}
 W(\phi)=\frac{6k}{\kappa^2}-u\phi^2\label{superpotential1}
\end{equation}
With such a choice,
the solution for the stabilizing scalar field ($\phi_0$) and the back-reacted metric are \cite{kribs,wolfe},
\begin{equation}
\phi(y)=\phi_{P} ~e^{-uy}\label{solution1}
\end{equation}
where $\phi_{P}$ is the solution at $y=0$. Similarly, the solution at
$y=r_0$ is $\phi_{T}$.
\begin{equation}
 A(y)=ky+\frac{\kappa^2 \phi_{P}^{2}}{12}e^{-2uy}\label{warpsolution2}
\end{equation}
The distance between two 3-brane $r_0$ can be stabilized
by matching the VEV of the stabilizing scalar field $\phi_0$ at 
$0$ (location of the Planck brane) and $r_0$ (location of the TeV brane)
to $\phi_P$ and $\phi_T$. This yields,
\begin{equation}
 e^{-ur_0}=\frac{\phi_T}{\phi_P}\label{brpotential}
\end{equation}
We slightly reparametrize the warp factor by using,
$l=\frac{\kappa\phi_P}{\sqrt{2}}$ and we obtain a new expression for 
warp factor which we will consider onwards.
\begin{equation}
 A(y)=ky+\frac{l^2}{6}e^{-2uy}\label{warpsolution3}
\end{equation}
\underline{Considering two possible cases:}\\
From the requirement of the desired warping from the Planck brane to TeV brane,
the contribution the warp factor will be,
\begin{equation}
 A(0)-A(r_0)=-37\label{warpingcondition}
\end{equation}
For $u>0$, $\phi_{P}^{2}-\phi_{T}^{2}>0$ which implies that the term $kr_0$
has the dominant contribution in the warp factor. This condition is justified
if $u/k$ is small i.e if we set $\phi_{P}/\phi_{T}=2.5$ and keep $l\lesssim1$, in turn $u/k$ becomes $1/40$.
For $u<0$ \cite{wolfe}, both the terms in warp factor
$\phi_{P}^{2}-\phi_{T}^{2}$ and $kr_0$ contribute to achieve the desired suppression from Planck to TeV brane.
For example, one could consider, $u/k=1/23$, $\phi_{T}/\phi_{P}=5$ while keeping $l\lesssim1$.\\
In this entire analysis, we take 
the 5-dimensional Planck scale $M$ and the 4-dimensional Planck scale $\overline{M}_{Pl}$ to be
of comparable magnitude,  as in the original RS model. 
% \begin{equation}
%  M^{2}_{Pl}=\frac{M^3}{k}\left[ (1-e^{-2kr_0})+\frac{2\kappa^2\phi_{P}^{2}}{12(1+\frac{u}{k})}
% (e^{-2(ur_0+kr_0)}-1)\right] \label{mplndm51}
% \end{equation}
As stated earlier, we consider the source of torsion to be the rank-2
anti-symmetric Kalb-Ramond field $B_{MN}$. Torsion can be identified
with the rank-3 antisymmetric field strength tensor $H_{MNL}$ which is
related to the KR field as,
\begin{equation}
 H_{MNL}=\partial_{[M}B_{NL]}\label{fldtensor}
\end{equation}
with each Latin and Greek indices running from $0$ to $4$
and $0$ to $3$ respectively.\\
After $S_1/Z_2$ orbifolding as in the original RS model, the
Kaluza-Klein decomposition for the KR field can be given as,
\begin{equation}
 B_{\mu\nu}(x,y)=\sum_{n=0}^{\infty}B^{n}_{\mu\nu}(x)
\frac{\chi^{n}(y)}{\sqrt{r_0}}\label{kkkr1}
\end{equation}
Substituting this in the 5-dimensional action and  integrating over the extra dimension,
the four dimensional effective action turns out to be:
\begin{eqnarray}
 S^{(4)}_{H}=\sum_{n=0}^{\infty}\int d^{4}x
[\eta^{\mu\alpha}\eta^{\nu\beta}\eta^{\lambda\gamma}H^{n}_{\mu\nu\lambda}
H^{n}_{\alpha\beta\gamma}\nonumber\\
+3m_{n}^{2}\eta^{\mu\alpha}\eta^{\nu\beta}B^{n}_{\mu\nu}
B^{n}_{\alpha\beta}]\label{efectiveaction4d}
\end{eqnarray}
provided the internal components of the KR field $\chi^{n}(y)$ satisfy,
\begin{equation}
 -\partial^{2}_{y}\chi^{n}(y)=m_{n}^{2}\chi^{n}(y)e^{2A(y)}\label{krdifeq1}
\end{equation}
along with  the orthonormality condition:
\begin{equation}
 \frac{1}{r_0}\int^{r_0}_{-r_0}e^{2A(y)}\chi^{m}(y)\chi^n(y)dy=\delta^{mn}\label{ortho1}
\end{equation}
Here $m_n$ represents the KK mass modes of KR field.

Let us now consider the massless mode of the KR field which is identified
with spacetime torsion. Eq.(\ref{krdifeq1}) yields the solutions
for the KR massless mode $\chi^{0}(y)$ for two distinct cases:\\
\underline{For $u>0$}, we obtain the wavefunction for the massless mode of KR field as,
\begin{equation}
\chi^{0}=\frac{\sqrt{kr_0}e^{-kr_0}}{\sqrt{1+\frac{l^2}{3(1-\frac{u}{k})}
\frac{\phi^{2}_{T}}{\phi_{P}^{2}}}}\label{wvsolution1}
\end{equation}
\underline{For $u<0$}, proceeding as above we obtain the
solution for the massless mode of KR field as,
\begin{equation}
 \chi^{0}=\frac{\sqrt{kr_0}e^{-kr_0}}{\sqrt{1+\frac{l^2}{3(1+\frac{u}{k})}
\frac{\phi^{2}_{T}}{\phi_{P}^{2}}}}\label{wvsolution2}
\end{equation}
 In both the cases the denominator in eq.(\ref{wvsolution1}) and
(\ref{wvsolution2}), is always $>1$. It was earlier shown in the
background of the original RS model \cite{ssgkr} that the torsion in
our 3-brane is diluted by a factor $\sqrt{kr_0}e^{-kr_0}$ in
comparison to the zero mode graviton. Here we find that due to the
back-reaction of the stabilizing field the space-time torsion is more
heavily suppressed than what was shown earlier in \cite{ssgkr}. This
establishes  the invisibility of torsion in our universe. 

We also estimate the coupling of the torsion field with standard model
fermions. \\ Assuming that the Standard Model matter fields are
confined to the brane at $y = r_0$, we evaluate the interaction of
torsion with spin $1/2$ brane fermions.  Following reference
\cite{ssgkr}, the interaction of the massless KR field with SM
fermions is therefore,
\begin{equation}
  \mathscr{L}_{\psi\overline{\psi}H^{0}}=-\frac{1}{M^{3/2}}\overline{\psi}
  \left[ i\gamma^{\mu}\sigma^{\nu\lambda}H^{0}_{\mu\nu\lambda}\frac{\chi^{0}(y)}
{\sqrt{r_0}}\right] \psi\label{couplingzero5}
\end{equation}
\underline{For $u>0$}:\\
Integrating over the extra dimensional part, we arrive
at the effective 4-D coupling of massless KR field with SM fermions.
\begin{equation}
 \mathscr{L}_{\psi\overline{\psi}H^{0}}=-i\overline{\psi}\gamma^{\mu}\sigma^{\nu\lambda}
\frac{e^{-kr_0}}{\overline{M}_{Pl}\sqrt{1+\frac{l^2}{3(1-\frac{u}{k})}
\frac{\phi^{2}_{T}}{\phi_{P}^{2}}}}H^{0}_{\mu\nu\lambda} \psi \label{KRcoupling1}
\end{equation}
\underline{For $u<0$}:
\begin{equation}
\mathscr{L}_{\psi\overline{\psi}H^{0}}=-i\overline{\psi}\gamma^{\mu}\sigma^{\nu\lambda}
\frac{e^{-kr_0}}{\overline{M}_{Pl}\sqrt{1+\frac{l^2}{3(1+\frac{u}{k})}
\frac{\phi^{2}_{T}}{\phi_{P}^{2}}}}H^{0}_{\mu\nu\lambda} \psi \label{KRcoupling2}
\end{equation}
We observe that the coupling is now suppressed by additional factors
for both $u$ greater or less than zero over the usual coupling
$\frac{e^{-kr_0}}{\overline{M}_{Pl}}$.  Thus the effect of the
backreaction of the stabilising bulk scalar field on the spacetime
geometry not only modifies the warp factor but also exhibits further
suppression in the interaction of torsion with fermionic fields on the
visible brane \cite{ssgkr}.  As is evident on binomially expanding the
right-hand sides of equations (23) and (24), the additional
suppression is due to the contribution of the bulk scalar field to
$T_{\mu\nu}$.  Further contributions of this kind from the bulk KR
fields will add to such suppression.  Therefore the back-reaction of
the KR field, if anything compared to the bulk scalar field, further
dilutes the effects of torsion in the SM brane.  This explains why the
search for any experimental detection of space-time torsion in our
universe continues to elude us, despite their presence alongwith pure
Einstien's gravity with equal strength in higher dimensions. The
miracle of warped compactification leads to the suppression of the
zero mode torsion while the couplings of the massless graviton with
brane fermions continue to remain $\sim M_P ^{-1} $.  A question that
still remains is: what happens to the massive modes of KR fields ?
Will they make their presence felt in our brane?  We finish our
discussion by addressing this issue.

For the massive modes, we consider two cases as before:\\
\underline{For $u>0$}, 
we reparametrize eq.(\ref{krdifeq1}) in terms of $z_n$, where
 $z_n=\frac{m_n}{k}e^{A(y)}$ and it is very clear that $\frac{l^2}{6}e^{-2uy}<1$
for any value of $y$ within $0$ to $r_0$ while we keep $l<1$.\\
Hence we obtain $e^{A(y)}\approx e^{ky}(1+\frac{l^2}{6}e^{-2uy})$,\\
and as a result eq.(\ref{krdifeq1})
reduces to,
\begin{equation}
 \left[ z_{n}^{2}\frac{d^2}{dz_{n}^{2}}+z_n\frac{d}{dz_n}+z_{n}^{2}\right] \chi^{n}=0\label{krdifeq2}
\end{equation}
The solution of the above equation is,
\begin{equation}
\chi^n=\frac{1}{N_n}\left[ J_0(z_n)+\alpha_n Y_0(z_n)\right] \label{solmassive1}
\end{equation}
Where $J_0(z_n)$ and $Y_0(z_n)$ are Bessel and Neumann function of the order $0$
and $\alpha_n$ and $m_n$ can be determined by applying the continuity condition
of the derivative of the wave function of massive KR at $y=0$ and $r_0$.
As explained in \cite{ssgkr}, applying the continuity condition at $y=0$
provides $\alpha_n<<1$ as, $e^{A(r_0)}>>1$. Here  $m_n$ are the KK mass modes on the TeV brane.
Applying the orthonormality condition eq.(\ref{ortho1})
to normalize $\chi^n$, we arrive at the final solution for 
$\chi^{n}$ as,
\begin{equation}
 \chi^n=\sqrt{kr_0}e^{-A(r_0)}J_0(z_n)\label{finalsolutionmsv}
\end{equation}
Proceeding similarly  as for the zeroth mode, we write the general expression for coupling
of massive KR fields with SM fermions on the TeV brane as
\begin{eqnarray}
\mathscr{L}_{\psi\overline{\psi}H^{n}}=-i\frac{e^{-A(r_0)}}{\overline{M}_{Pl}}
\overline{\psi}\gamma^{\mu}\sigma^{\nu\lambda}\\ \nonumber
\sum_{n=1}^{\infty}H^{n}_{\mu\nu\lambda}J_0(x_n)\psi\label{couplingmsv2}
\end{eqnarray}

\underline{For $u<0$}, one can perform a similar analysis. There, too,
we can achieve the same suppression for massive KR mode wave functions
and their coupling with SM fermions.

Thus a warped extra dimensional theory like the RS scenario is
successful in explaining the suppression of torsion in our
4-dimensional universe. Apart from a large exponential suppression by the warp factor, the back reactions caused by the energy density
of the stabilising field in the bulk further accentuates such
suppression. A similar fate greets the massive modes of the KR field,
when projections on the visible brane are taken. 

{\bf Acklnolowledgements:} The work of BM has been partially supported
by funding available from the Department of Atomic Energy, Government of
India, for the Regional Centre for Acceleratror-based Particle Physics (RECAPP) 
at Harish-Chandra Research Institute. BM also thanks the Theoretical Physics
Depertment, Indian Associatoin for the Cultivation of Science, for hospitality
during the early stage of this work, while SSG acknowledges the hospitality of
RECAPP while the project was being completed.

\end{document}